\documentstyle[amsfonts,eqsecnum,aps]{revtex}
\begin{document}

\preprint{SPhT-95-152}

\title{Distribution of roots of random real generalized polynomials}
\author{G.~Andrei Mezincescu$^{1,2}$, Daniel Bessis$^{1,3}$, 
Jean-Daniel Fournier$^{1,4}$, \\Giorgio Mantica$^{1,5}$   
and Francisc D. Aaron$^6$}
\address{$^1$ Service de Physique Th\'eorique, C.E. Saclay, 
F-91191 Gif-sur-Yvette Cedex, France\\ 
$^2$ Institutul de Fizica \c si Tehnologia Materialelor, C.P. MG-7, 
Bucure\c sti -- M\u agurele, Rom\^ania\\ 
$^3$ CTSPS, Clark-Atlanta University, Atlanta, GA 30314\\
$^4$ Laboratoire Cassini, Observatoire de Nice, BP 229, F-06304 Nice Cedex 4, 
France\\
$^5$ Universita di Como, Via Lucini 3, I-22100 Como, Italia\\
$^6$ Facultatea de Fizic\u a, Universitatea 
Bucure\c sti, C.P. MG-11, Bucure\c sti -- M\u agurele, Rom\^ania}
\date{December 20, 1995;~~ Revised May 29, 1996}  
\maketitle 
\vskip 2em

{\bf Abstract.} 
The average density of zeros for monic generalized polynomials, 
$P_n(z)=\phi(z)+\sum_{k=1}^nc_kf_k(z)$, with real holomorphic 
$\phi ,f_k$ and real Gaussian coefficients is expressed in terms 
of correlation functions of the values of the polynomial and its 
derivative.  We obtain compact expressions for both the regular 
component (generated by the complex roots) and the singular one
(real roots) of the average density of roots. The density of the 
regular component goes to zero in the vicinity of the real axis 
like $|\hbox{\rm Im}\,z|$.  
We present the low and high disorder asymptotic behaviors. 
Then we particularize to the large $n$ limit of the average density 
of complex roots of monic algebraic polynomials of the 
form $P_n(z) = z^n +\sum_{k=1}^{n}c_kz^{n-k}$ with real 
independent, identically distributed Gaussian coefficients 
having zero mean and dispersion $\delta = \frac 1{\sqrt{n\lambda}}$. 
The average density tends to a simple, {\em universal} 
function of $\xi={2n}{\log |z|}$ and $\lambda$ in the domain 
$\xi\coth \frac{\xi}{2}\ll n|\sin \arg (z)|$
where nearly all the roots are located for large $n$. 

%

\def\E{{\Bbb E}}               
\def\R{{\Bbb R}}               
\def\P{{\Bbb P}}               
\def\Z{{\Bbb Z}}               
\def\N{{\Bbb N}}               
\def\Q{{\Bbb Q}}               
\def\C{{\Bbb C}}               

\def\d{\delta}
\def\eps{\epsilon}
\def\o{\omega}
\def\vp{\theta}

\def\calE{{\cal E}}
\def\calD{{\cal D}}
\def\calH{{\cal H}}
\def\calQ{{\cal Q}}
\def\calO{{\cal O}}
\def\calN{{\cal N}}
\def\calK{{\cal K}}
\def\calR{{\cal R}}
\def\calP{{\cal P}}

\def\Re{\hbox{\rm Re}}
\def\Im{\hbox{\rm Im}}
\def\ci{{\scriptstyle {\circ} }\displaystyle }
\def\rmi{{\rm{i}}}
\def\rme{{\rm{e}}}  
\def\rmd{{\rm{d}}}  
\def\rmK{{\sf{K}}}  
\def\rmM{{\sf{M}}}
\def\rmT{{\sf{T}}}  
\def\sn{\sum_{k=1}^n} 
\def\szeron{\sum_{k=0}^{n-1}} 
\def\s2{\sum_{\alpha=1}^2} 
\def\Cn{C_n(r^2,2\varphi)} 
\def\Cnzero{C_n(r^2,0)} 
\def\Sn{S_n(r^2,2\varphi)} 

\def\a{\bbox{\rm a}}  
\def\c{\bbox{\rm c}}  
\def\f{\bbox{\rm f}}
\def\h{\bbox{\rm h}}
\def\g{\bbox{\rm g}}

\def\df{\dot f}
\def\dF{\dot F}
\def\dP{\dot P}
\def\dF{\dot F}
\def\dfi{\dot \phi}           
\def\dpsi{\dot \psi}
\def\dbf{\dot{\f}}


\begin{section}{Introduction}\label{intro}

Let $P_n$ be a (monic) algebraic polynomial of degree $n$, 
\begin{equation}
P_n(z)= z^n + a_1z^{n-1}+\cdots +a_n.\label{1.1}
\end{equation}
The roots of $P_n$ are (algebraic) functions of its coefficients 
$a_1,a_2,\cdots,a_n$. 
If the coefficients are random variables, then the roots of the 
polynomial will also be random variables. The study of the 
distribution of the roots of random polynomials began with the 
investigation by Bloch and Polya \cite{BP32}, continued by 
Littlewood and Offord \cite{LO38,LO39}, Kac \cite{Kac59} and 
many others (for reviews see \cite{B-RS86,EK95}) of the number and 
distribution law of the real roots of random polynomials 
with real coefficients.

Most investigations of the distribution of zeros 
for random algebraic polynomials have dealt either with the 
initial problem -- the real zeros of real polynomials -- or 
with the complex zeros of complex polynomials. In the latter 
case, one is helped by the fact that it is possible
to transform the joint distribution function of the coefficients 
into the joint distribution function of the roots explicitly 
using the relations between the coefficients of an algebraic 
polynomial and its roots.\footnote{Which follow by comparing 
the polynomial $P_n(z) = \sum_{k=0}^na_kz^{n-k}$ with its 
roots expansion $P_n(z) = a_0\prod_{j=1}^n (z-z_j)$. The Jacobi 
determinant for this transformation was computed by Girschick 
and Hammersley \cite{girshick42,hamm56,BBL}.}   
The average distribution of roots for homogeneous 
algebraic polynomials,  
\[P_{n-1}(z)=\sum_{j=0}^{n-1}c_jz^j,\] 
with real normal Gaussian coefficients $c_j$, was recently 
investigated by Shepp and Vanderbei\footnote{We thank an 
anonymous referee for bringing this paper to our attention} 
\cite{SV95}.  They obtain a generalized Kac-Rice formula for 
the density  of complex roots and explore its large $n$ limit.  
The fraction of the expected number of roots contained in an 
angular sector $S(\theta_1,\theta_2)$ which does not intersect 
the real axis\footnote{The analogous result for complex 
coefficients was obtained by Shparo and Shur\cite{ss62}.} 
tends to $|\theta_2-\theta_1|/2\pi$.
Most roots are concentrated in a small annulus 
(with width $\sim n^{-1}$) near the unit circle. 
${\cal N}(R)$, the expected number of roots in a ball of 
radius $R$, satisfies 
\begin{equation}
\lim_{n\to\infty}\frac{1}{n}{\cal N}(\rme^{\frac{s}{2n}}) =
\frac {1}{1-\rme^{-s}}-\frac 1s .\label{sv}
\end{equation}
In a recent preprint by Ibragimov and Zeitouni  
this result is generalized to coefficient distributions 
belonging to the domain of attraction of a $\alpha$-stable
law \cite{IZ96}.

In this paper we study the distribution of the roots for more 
general random polynomials with {\it real} coefficients. 
The initial motivation for the present work was given by some 
interesting properties of the roots of the Szeg\"o orthogonal 
polynomials \cite{szego} related to the Wiener transfer function 
\cite{Wi49}. Szeg\"o polynomials are orthogonal polynomials 
in the non negative measure on the unit circle of the complex plane 
$\rmd\mu(\vp)$ generated by a non-negative T\"oplitz form 
$T(k-l)\,;k,l=0,1,\ldots$. Here 
\begin{equation}
T(k)=\int_{-\pi}^{\pi}\rmd\mu(\vp)\rme^{-\rmi k\vp} \label{1.T}
\end{equation}
are the measure's moments. In the case of the Wiener transfer function 
the moments are equal to the autocorrelation function of a discrete 
finite real signal sample ($X(p)=0;~p<0 ~\hbox{or}~p\ge N$): 
\begin{equation}
T(k)=F_N(k) = \sum_{p=0}^{N-1} X(p)X(p+k) \label{1.TN} 
\end{equation}
Let the signal be the sum of a useful signal, consisting of several 
harmonic components with frequencies $\omega_\nu$, and noise. Then there 
is strong numerical and some analytic evidence that for signal-to-noise 
ratios that are not too low, the roots of the Szeg\"o polynomials of order 
$n$ with $1\ll n\ll N$  break into two groups. The first one converges 
rapidly to the unit circle at the points $\rme^{\rmi\omega_\nu}$, 
\cite{pakula87,JNS90,JN91}. The other group, which also converges to the unit 
circle, but more slowly, is nearly equispaced -- resembling an one-dimensional 
crystal. It has universal statistical properties even when the harmonic 
components of the signal are absent \cite{FMMB,FMMB94}. 

If the useful signal is absent, the Szeg\"o polynomial of  
order $n$ is well approximated by (\ref{1.1}) with $a_k=N^{-\frac 12}c_k$ 
with approximately Gaussian coefficients $c_k$ if $1\ll n\ll N$ \cite{FMMB}.  
In the presence of a useful signal, the polynomials have a more 
complicated form, which can be obtained by substituting more general 
polynomials for the monomials $z^k$ in (\ref{1.1}). Thus, we are led to 
considering {\em generalized random polynomials}.

Let 
\begin{equation} 
\phi(z) ,\  f_k(z),\ k=1,2,\ldots,n~~ ,~n\in{\N} 
\label{1.10}\end{equation}
be holomorphic and linearly independent functions of the 
complex variable $z$ in a domain of the complex plane, 
$z = x+\rmi y\in D\subset{\C}$. Let $\c (\o )$ be a 
random $n$-dimensional vector with components 
$c_k(\o ),\ k=1,2,\ldots,n\ ,~n\in{\N}$. Here $\o\in\Omega$ 
where $\Omega$ is a probability space. 

We define random generalized {\em monic} polynomials 
of degree $n$ by
\begin{equation}
P_n(z)=\phi (z)+\sum_{1\le k \le n} c_kf_k(z). \label{1.11}
\end{equation}
{\em Homogeneous} polynomials (of degree $n-1$) correspond to setting 
$\phi\equiv 0$. Alternatively, a monic polynomial of degree $n$ 
can be regarded as a homogeneous one, having a singular distribution 
for the coefficient $c_{n+1}$ ($\delta(c_{n+1}-1)$). 
Setting $\phi(z) = z^n, f_k(z) = z^{n-k} ,\ k=1,2,\ldots,n\ ,n\in{\N}$
we obtain the random monic algebraic polynomials, (\ref{1.1}). 
Trigonometric, hyperbolic and other types 
of random polynomials \cite{B-RS86} can be obtained by 
suitably defining  $\phi$ and $f_k$.  In the following we 
will often omit the qualifier {\it generalized} and call 
the objects defined by Eq.(\ref{1.11}) simply polynomials. 

Let us note that any non-random affine transformation of the 
random vector of the coefficients, 
\begin{equation}
c_k \to a_k+\sum\rmK_{km}\tilde{c}_m\label{aff.1}
\end{equation}
where $\a$ is a constant vector and $\rmK$ a non singular 
matrix, transforms the polynomial $P_n$ into one of the same 
form (\ref{1.11}) but with the basis set (\ref{1.10}) replaced by
\begin{equation} 
\tilde\phi=\phi +\sum_{m}a_mf_m;,~ \  \tilde{f}_k=
\sum_{m}\rmK_{mk}f_m.  \label{aff.2}\end{equation}
Thus, by redefining the basis set, we can always consider that 
the mathematical expectation of the polynomial is equal to its 
deterministic part, $\phi$. 

Let $F(z;\o )$ be a random holomorphic function of $z\in D\subset\C$, 
{\it i.e.} a family of functions  $F(z;\o )$ indexed by $\o\in\Omega$ 
which are almost surely holomorphic for $z$ in the domain $D$. Let 
\begin{equation} 
z_r(\o ) = x_r(\o ) + iy_r(\o ),\ r = 1,\ldots  \label{1.2}
\end{equation}
be the solutions of the equation
\begin{equation}
F(z;\o )=0. \label{1.3}
\end{equation}
The zeros of $F$ are random variables. In each compact subdomain 
${D}_1\subset D$, there may be only a finite number $N(D_1;\o )$, 
of zeros for each realization $\omega\in\Omega$, since the accumulation 
points of zeros cannot lie inside the domain of holomorphy.

The density of zeros of the random function $F$ is the 
random distribution on ${\R^2}$:
\begin{eqnarray} 
\rho (x,y;\o ) &=& \sum_r\delta (x-x_r(\o ))\delta (y-y_r(\o ))\nonumber\\     
&=& \sum_r \delta^{(2)}(z-z_r(\o )),\label{1.4}
\end{eqnarray}
where $\delta $ and $\delta^{(2)}$ are, respectively, the 
Dirac distribution on $\R$ and on $\R^2$. Using the definition 
of the Dirac $\d$ we may rewrite (\ref{1.4}) as 
\begin{equation}
\rho (x,y;\o )= \left\vert\dF (z;\o )\right\vert^2 \d^{(2)}
\big (F(z;\o )\big ), \label{1.5}\end{equation}
where the Cauchy-Riemann conditions were used to calculate the Jacobian. 
The compact notation 
\begin{equation}\dF = \frac{\rmd F}{\rmd z} \end{equation}
for the derivative of $F$ will be used throughout this paper.

The expected (average) density of zeros of $F$ is obtained by
averaging over the realizations:
\begin{equation} 
{\cal D}(x,y) = \E \left\{ \rho (x,y; \ci ) \right\} .  \label{1.6}
\end{equation}
In a similar way, one may define the two-point correlation function of 
the zeros by
\begin{equation} 
{\cal D}_2(x_1,y_1;x_2,y_2)=\E\left\{\rho (x_1,y_1;\ci ) 
\rho (x_2,y_2;\ci )\right\}  \label{1.7}
\end{equation}
and higher, $m$-point, correlation functions for the roots. 

Substituting (\ref{1.5}) into (\ref{1.6}) and introducing 
the joint distribution function of the values of the function and 
its derivative at the point $z=x+\rmi y$
\begin{equation}
\calP (\alpha,\tilde{\alpha};z)=
\E\left\{\d^{(2)}\big (\alpha-F(z;\ci )\big )
\d^{(2)}\big (\tilde{\alpha}-\dF (z;\ci )\big )\right\}, 
\label{1.8}\end{equation}
we see that the average density of roots at the point 
$z=x+\rmi y$ is given by the Kac-Rice \cite{Kac59,B-RS86,Rice45} 
formula
\begin{eqnarray}
{\cal D}(x,y)&=&\E\{\vert\dF (z;\ci )\vert^2 \d^{(2)}\big (F(z;\ci )\big )\}
\nonumber\\&=&\int \vert\tilde{\alpha}\vert^2
\calP (0,\tilde{\alpha};z)\rmd^{(2)}\tilde{\alpha}.
\label{1.9}
\end{eqnarray}
in terms of $\calP (\alpha,\tilde{\alpha};z)$. 
Similar formulas, involving only the joint distribution function 
of the values of the function and its derivative at the selected 
points may be written for the $m$-point correlation functions of 
the roots such as (\ref{1.7}). We will study the average 
distribution of roots, Eq.(\ref{1.6}), in the case when $F(z,\o )$ 
is a generalized random polynomial, (\ref{1.11}), with Gaussian 
coefficients. 

If the components of $\c (\o )$ in (\ref{1.11}) 
are Gaussian (real or complex) random variables, the joint 
distribution function of the values of the polynomial  
and its derivative at some point $z$ (\ref{1.8}) will also be 
also a Gaussian distribution which is determined by the 
correlation functions (covariance matrix) of the values of the 
polynomial and its derivative at that point. 
This means that we may compute explicitly the 
integral in Eq.(\ref{1.9}) for the average density of roots 
in the Gaussian case and the corresponding expressions 
for the $m$-point correlation functions of the roots. 

In Section \ref{2} we will obtain the average density of 
roots, Eq.(\ref{1.6}), for {\it real} Gaussian generalized 
monic polynomials, when the basis functions in (\ref{1.10}) 
are of {\em real} type
\begin{equation}
\left[\phi(z)\right]^*=\phi(z^*),~~\left[f_k(z)\right]^*=f_k(z^*).
\label{1.12}\end{equation}
Here $~^*~$ denotes complex conjugation. The coefficients 
$c_k$ are independent, identically distributed ({\it (iid)}) 
real Gaussian random variables with zero expectation value. 
The assumption of a joint Gaussian distribution is important, 
while the restriction to the {\it iid} case is inconsequential. 
Indeed, by a suitable linear mapping of type (\ref{aff.1}) 
any finite Gaussian distribution may be mapped onto the standard 
one ({\em iid} with zero average and unit dispersion). 
By the above remark this leads to a redefinition (\ref{aff.2}) 
of the basis set. 

We will obtain a general formula giving the expected density 
of complex roots at points with $\Im (z)\ne 0$. Due to the 
reality condition, (\ref{1.12}), there is also a singular 
component of the expected density of roots located on the 
real axis. We will obtain a generalization of the Kac-Rice 
formula for it. For homogeneous polynomials with non-central  
Gaussian distributed coefficients we recover a result by 
Edelman and Kostlan [6], who also pointed out that taking 
a singular correlation matrix limit for that case will  
yield the density of real roots in the monic case. 
We will show that the density of complex 
roots tends to zero, like $|\Im (z)|$, in the vicinity of 
the real axis.  In the high randomness limit (large 
dispersion $\delta$ of the Gaussian distribution) the 
expected density of roots for monic polynomials approaches, 
obviously, that for the homogeneous ones ($\phi\equiv 0$).
In the weak randomness limit, when the dispersion $\delta$ 
goes to zero, the roots concentrate near the zeros of the 
deterministic part $\phi$. Near simple zeros of $\phi$ the 
distribution tends to a Gaussian one. Near zeros of $\phi$ with 
multiplicity $k>1$, it has generically $2k$ maxima at a distance 
$\sim \delta^{\frac 1k}$ from the zero's position. An interesting 
situation arises when the multiplicity of the zero is large.

Since only the joint distribution function of the polynomial 
and its derivative is assumed to be Gaussian, the results of 
Section \ref{2} could also be relevant to cases 
when the coefficients are non-Gaussian but one may prove a 
limit theorem for the joint distribution function. 

In Section \ref{3} we will apply the general results of 
Section \ref{2} to the density of roots of monic algebraic 
random polynomials, (\ref{1.1}), with {\it iid} coefficients 
having a real Gaussian distribution with zero average 
and dispersion $\delta = N^{-\frac 12}$ in the large $n$ and 
$N$ limit. In this case the deterministic part has a real 
zero of multiplicity $n$ at the origin. 
As mentioned above, in \cite{FMMB} the Szeg\"o 
polynomials associated to the Wiener transfer function for 
pure noise were shown to have this form if the sample 
length is $N \gg n\gg 1$. 

Defining a new rescaled coordinate $\xi = {2n} \ln |z|$  
and rescaling also the dispersion, $\delta^{-1}=N=n\lambda$, 
we obtain a simple asymptotic expression for the average 
density of roots valid for large $n$ at points which satisfy   
\begin{equation}
\xi\coth \frac{\xi}{2}\ll n|\sin \arg (z)|.\label{domain}
\end{equation}
This is a small neighborhood of the unit circle which does not 
intersect the real axis and it contains nearly all the zeros. 
To leading order, the average density in this domain does 
not depend on $\arg (z)$. It approaches $n^2$ times a universal 
function of $\xi$ and $\lambda$. For $\sin\theta\ne 0$,
\begin{equation}
\lim_{n\to\infty}\frac 1{n^{2}}
{\cal D}({\rm e}^{\frac{\xi}{2n}}, \vp) 
= -\frac{1}{\pi}\frac{\rmd}{\rmd\xi} \Biggl\{
\left (\frac{1}{\xi} - \frac{1}{{\rm e}^\xi - 1}\right )
\exp\left [{-\lambda\frac{\xi}{1 - {\rm e}^{-\xi}}}\right ]
\Biggr\}. \label{fin.1}
\end{equation} 

In the case of homogeneous polynomials, which corresponds to 
$\lambda=0$, (\ref{fin.1}) is equivalent to (\ref{sv}),  
the result of Shepp and Vanderbei \cite{SV95}.   

We will finally remark that the asymptotic distribution 
of complex roots for polynomials with {\em real} Gaussian 
coefficients, (\ref{fin.1}), which do not have the rotational 
symmetry of the polynomials with {\em complex} coefficients, 
nevertheless coincides with the rotationally invariant one 
for the latter, which can be readily obtained using the methods we use.
For the complex homogeneous case ($\lambda=0$), the asymptotic   
estimate (\ref{sv}) is due to Arnold \cite{arnold66,B-RS86}. 

A part of the calculations for the algebraic polynomials will 
be presented in more detail in the Appendix.
Before proceeding further, we will define in the next section  
some notations that will allow us to write in a compact way 
the joint distribution function (\ref{1.8}) and our results  
when the random function is a polynomial with Gaussian coefficients.

\vskip 2em
{\bf Acknowledgments} GAM, DB and JDF acknowledge the 
support of the European Science Foundation for 
participation at the Como Workshop {\em Classical Mechanics 
Methods in Quantum Mechanics}; 
GAM and GM thank Alfred Msezane and Carlos Handy for hospitality 
at CTSPS, Clark Atlanta University, where parts of this work were done. 
JDF and GM acknowledge support from the {\it Programme 
Galilee de collaboration scientifique France - Italie}. 
\end{section}


\begin{section}{A digression on notations}
\label{not}

For any holomorphic function, $h$, we will write  
\begin{equation}
{dh\over{dz}} = \dot{h}. \label{2.1}
\end{equation}
for its derivative. The real and imaginary parts of a 
complex number or function $h$ will be denoted by the 
subscripts {\em 1} and {\em 2} respectively.
\begin{equation} 
         h_1 = \Re (h), \ h_2 = \Im (h).   \label{2.2}
\end{equation}        

As remarked in the Introduction, for a given realization 
the coefficients define a point in ${\R^n}$, while the 
basis set of functions, $\phi$ and $f_k,~k=1\ldots n$,  
maps $\C\sim R^2$ in $\C\times\C^n~ \sim \R^2\times\R^{2n}$. 
The values of the polynomial $P_n(z)$, Eq.(\ref{1.11}), and 
its derivative $\dot P_n(z)$ are affine mappings (with complex 
coefficients) of the real $n$-vector of the coefficients  
into $\C\sim R^2$. Let us introduce more compact notations 
for the various vector structures we are dealing with. 

We will use the {\it same} lower case {\it Greek} letter 
({\it e.g.}~ $\phi$) for {\it both} the {\it complex scalar} 
$\phi \in {\C}$ and the  {\it real 2-vector} $\phi$ whose 
components $(\phi_1,\phi_2)\in\R^2$ are the real and 
imaginary parts of the complex scalar $\phi$.  

Boldface letters denote real $n$-vectors 
$\R^n\ni\f = (f_1,f_2,\ldots ,f_n)$.  When it does not lead to 
ambiguities, we might abuse the notation to denote a complex 
$n$-vector -- like the vector made of the basis set of functions 
$f_k(z),~k=1,2\ldots ,n$~ -- by $\f(z)$. In such cases, the real 
and imaginary parts are $\f_1(x,y), ~\f_2(x,y)$ and we may write
$\f(z) = \f_1(x,y) + \rmi\f_2(x,y)$.

{\it Latin} letters, $f,~g$, will be used for {\it real} 
$2n$-vectors with components $f_{k\alpha}$, 
$g_{k\alpha},~ k = 1,2,\ldots ,n,\ \alpha=1,2$. 
We use {\it the same letter} in the typefaces mentioned above 
for related objects, like 
$\f(z),~\bigl(\f_1(x,y), ~\f_2(x,y)\bigr),~f(x,y)$ 
for the basis set of functions at point $z$ considered 
as a complex $n$-vector made of two real $n$-vectors or as 
a real $2n$-vector.

Greek subscripts always run over the set $\{ 1,2 \}$ while 
the Latin ones over  $\{ 1,2,\ldots ,n\}$.

For a Gaussian distribution of the coefficients, 
the joint distribution function of the values of the random 
polynomial $P_n(z)$, Eq.(\ref{1.11}), and its derivative 
$\dot P_n(z)$ is determined by their correlation matrix.  
The calculation of averages can be done using 
Wick's theorem, replacing the products of coefficients by 
their expectation value. This reduces further to a contraction 
over the Latin indices if the Gaussian process is a direct sum of 
normal ones.\footnote{As noted in the introduction this is 
always possible choosing a suitable (\ref{aff.1}).}
The compact notations defined in the rest of 
this section will help us with the necessary bookkeeping.

We use {\em bold square brackets} $\biglb[\ci ,\ci\bigrb]$ 
for the contractions over the Greek 
indices and {\em bold round brackets\ } $\biglb(\ci ,\ci\bigrb)$ for 
the contractions over the Latin ones. Using these conventions,
\begin{equation} 
\biglb[\phi ,\psi\bigrb]=\sum_{\alpha = 1}^2 
\phi_\alpha\psi_\alpha, \label{2.3}
\end{equation}
will be {\it the real scalar product} of the $2$-vectors $\phi$ and $\psi$; 
$\biglb[\phi ,f\bigrb]$  will denote the real $n$-vector with components
\begin{equation} 
{\biglb[\phi , f\bigrb]}_k =  
\sum_{\alpha = 1} ^2 \phi_\alpha f_{k\alpha},  \label{2.4}
\end{equation}
obtained by contacting the direct product between the 2-vector $\phi$ and
the 2n-vector $f$ over the Greek indices. 
In an analogous way, $\biglb[f,g\bigrb]$ will be the second order tensor 
obtained by contracting over the Greek indices the direct product of the
2n-vectors $f$ and $g$. Its components form the $n\times n$ matrix
\begin{equation} 
{\biglb[f,g\bigrb]}_{jk} = 
\sum_{\alpha = 1}^2f_{j\alpha}g_{k\alpha},\label{2.5}
\end{equation}

In  a similar way, the scalar product between the real n-vectors $\f$  
and  $\g$ is 
\begin{equation} 
\biglb(\f ,\g\bigrb) =\sum_{k=1}^nf_k g_k,\label{2.6}
\end{equation}
while
\begin{equation} 
\biglb(f,g\bigrb)_{\alpha\beta} =\sn f_{k\alpha}g_{k\beta},\label{2.7}
\end{equation}  
are the elements of the $2\times 2$ (real) matrix $\biglb(f,g\biglb)$,
while
\begin{equation} 
\biglb(\c,f\bigrb)_{\alpha} =\sn c_{k}f_{k\alpha},\label{2.8}
\end{equation}  
are the components of a $2$-vector.

Finally, we use matrix notations like $\biglb(f,g\bigrb)\psi$ 
for the product of the $2\times 2$ matrix $\biglb(f,g\bigrb)$ 
and the 2-vector $\psi$, 
\begin{equation}
\bigl(\biglb(f,g\bigrb)\psi \bigr)_\alpha = 
\sum_{\beta =0}^2\biglb(f,g\bigrb)_{\alpha\beta}\psi_\beta.
\end{equation}  

\end{section}


\begin{section} {Average density of roots for random Gaussian 
generalized monic polynomials}
\label{2}
In this section we consider monic holomorphic polynomials of 
type (\ref{1.11}) with coefficients distributed according to 
a Gaussian law. As we remarked in the Introduction, the case 
of homogeneous (non monic) polynomials may be obtained by 
setting $\phi \equiv 0$. There we also noted that it suffices 
to consider only the case of {\it iid} normal Gaussian coefficients 
(with zero expectation and unit dispersion):
\begin{eqnarray}                                 
\E\{c_k\} =& 0,~~&  \nonumber \\
&~~~~& j,k = 1,2,~  \ldots , n. \label{3.1} \\ 
\E\{c_kc_j\} =& \delta_{kj},   \nonumber
\end{eqnarray}

The calculation of the average density of roots, their two-point 
correlation function  $\calD,\ \calD_2$  and even higher correlation 
functions $\calD_m$ may be done in closed form since all the 
integrals will be Gaussian, although the formulae will tend to 
become rather encumbering with increasing $m$.        

\subsection{Average density of complex roots}\label{complexr}
The starting point will be the formula (\ref{1.9}) for the 
average density of roots for the polynomial 
\begin{equation}
P(z)=\phi (z)+F(z),\label{3.2}
\end{equation}
where $\phi$ is the deterministic (non random) part and 
\[F(z)=\sum_{k=1}^nc_kf_k(z)=\biglb(\c,f(z)\bigrb).\]
Here $\c$ is the real vector of the Gaussian coefficients, and 
we use the bold round bracket notation (\ref{2.8}) introduced 
in the previous section. 

For the sake of clarity of the main points of the calculation, 
we will first illustrate our approach in the case of homogeneous 
(non monic) polynomials,  {\em i.e.}
\[\phi\equiv 0,\] 
obtaining a generalization of the results derived by Shepp and 
Vanderbei \cite{SV95} in the algebraic case ($f_j(z)=z^{j-1}$). 
Subsequently we will obtain the average density of complex  
roots in general case of monic polynomials. 

In the homogeneous case the joint 
distribution function for the polynomial and its derivative, 
$P$ and $\dot P$, at the point $z=x+\rmi y$ coincides with the 
one for $F$ and $\dot F$ at the same point. Then,
\begin{equation}
{\cal D}=\int\biglb[\tilde{\alpha},\tilde{\alpha}\bigrb]
{\cal P}(0,\tilde{\alpha})\rmd^{(2)}\tilde{\alpha},
\label{3.3}\end{equation}
Here we used the bold square bracket notation (\ref{2.3}) 
for the contractions introduced in section \ref{not} to 
rewrite $|\tilde{\alpha}|^2$ as $\biglb[\tilde{\alpha},\tilde{\alpha}\bigrb]$. 

Now, $\cal P$ will be a Gaussian distribution determined by 
$\Delta (z)$, the $4\times 4$ correlation matrix of $F(z)$ 
and $\dot{F}(z)$. Its elements are the expectation values of 
products of elements of the $4$-vector 
$\left(\Re (F),\Im(F),\Re(\dot{F}),\Im(\dot{F})\right)^T=
\left(F_1,F_2,\dot{F}_1,\dot{F}_2\right)^T$. 
For example:
\[\Delta_{14}(z)=\E\{\Re(F(z))\ \Im(\dot{F}(z))\}=
\sum_{k,j}f_{k1}(z)\ \df_{j2}(z)\E\{c_kc_j\}. \]
Using (\ref{3.1}) this reduces to
\[\Delta_{14}(z)=\sum_kf_{k1}(z))\ \df_{k2}(z). \]
We may use the bold round bracket notation (\ref{2.7}) to 
write the 4$\times$4 matrix $\Delta (z)$ as a 2$\times$2 
block matrix whose elements are 2$\times$2 matrices
\begin{equation}
\Delta (z)=\pmatrix{
\biglb(f(z),f(z)\bigrb)&\biglb(f(z),\df(z)\bigrb)\cr
\biglb(\df(z),f(z)\bigrb)&\biglb(\df(z),\df(z)\bigrb)
}.
\label{3.4}\end{equation}

As a correlation matrix $\Delta (z)$ is symmetric and non-negative. 
The same is true for the diagonal blocks in (\ref{3.4}). Since 
the functions $f_k(z)$ are linearly independent and of real type, 
(\ref{1.12}), $\Delta (z)$ will be generically non singular off 
the real axis. For $\Im (z)\ne 0$, we may then write the joint 
distribution function of $F$ and $\dF$
\begin{equation}
{\cal P}(\alpha ,\tilde{\alpha};z)=\frac 1{(2\pi)^2\sqrt{\det\,\Delta(z)}}
\exp \biggl [-\frac 12 (\alpha ^T,\tilde{\alpha}^T)
\Delta^{-1}(z)\left ({\alpha \atop\tilde{\alpha}}\right )\biggr ].
\label{3.5}\end{equation}
Here $\alpha =\pmatrix{\alpha _1\cr \alpha _2}$ and 
$\tilde{\alpha}=\pmatrix{\tilde{\alpha}_1\cr\tilde{\alpha}_2}$ 
are 2-component vectors and we use $^T$ to denote 
the transposed. 

For $\Im (z)\ne 0$, where the rank of $\Delta$ is equal to four, 
we may introduce the block-Cholesky decomposition of the positive 
matrix $\Delta$ into block-triangular 2$\times$2 factors:
\begin{equation}
\Delta=\rmK^T\rmK,
\label{3.6}\end{equation}
where the upper block triangular matrix 
\begin{equation}
\rmK=\pmatrix{\biglb(f,f\bigrb)^{\frac 12}&
\biglb(f,f\bigrb)^{-\frac 12}\biglb(f,\df\bigrb)\cr
0&\left [\biglb(\df,\df\bigrb)_\perp\right ]^{\frac 12}}.
\label{3.7}\end{equation}
Here we introduced the notation
\begin{equation}
\biglb(\df,\df\bigrb)_\perp = 
\biglb(\df -\rmT_\perp f,\df -\rmT_\perp f\bigrb)
=\biglb(\df,\df\bigrb)-
\biglb(\df,f\bigrb)\biglb(f,f\bigrb)^{-1}\biglb(f,\df\bigrb),
\label{3.8}\end{equation}
with the matrix 
\begin{equation}
\rmT_\perp = \biglb(\df,f\bigrb)\biglb(f,f\bigrb)^{-1}.
\label{3.9}\end{equation}
The square root of a matrix, $\rmM$, is defined by the well 
known relation
\[
\rmM^{\frac 12}=\frac 2\pi\int_0^\infty\frac{\rmd t}{1+t^2\rmM^{-1}},
\]
which is true as long as $\rmM$ has no eigenvalues on the 
interval $(-\infty,0]$. For brevity's sake we also dropped the 
dependence on $z=x+\rmi y$, which we will continue to do 
whenever this does not lead to ambiguities.

If the matrix $\Delta$ is strictly positive, its diagonal block 
$\biglb(f,f\bigrb)$ has the same property. Then,
\[\Delta^{-1}=\rmK^{-1}\left (\rmK^{-1}\right )^T \] where
\begin{equation}
\rmK^{-1}=\pmatrix{\biglb(f,f\bigrb)^{-\frac 12}&
-\biglb(f,f\bigrb)^{-1}\biglb(f,\df\bigrb)
\left [\biglb(\df,\df\bigrb)_\perp\right ]^{-\frac 12}\cr
0&\left [\biglb(\df,\df\bigrb)_\perp\right ]^{-\frac 12}
}.
\label{3.10}\end{equation}
The matrix $\biglb(\df,\df\bigrb)_\perp$ was defined above, (\ref{3.8}).

Substituting (\ref{3.5}) into (\ref{3.3}) we may rewrite (\ref{3.3}) as
\begin{equation}
{\cal D}=\frac 1{(2\pi )^2\sqrt{\det\,\Delta}}
\int\biglb[\tilde{\alpha},\tilde{\alpha}\bigrb]\exp \left\{-\frac 12 
\bigglb[\tilde{\alpha},\big (\biglb(\df,\df\bigrb)_\perp\big )^{-1}
\tilde{\alpha}\biggrb]\right\}\rmd^{(2)}\tilde{\alpha}.
\label{3.11}\end{equation}

This Gaussian integral is readily evaluated yielding
\[
{\cal D}=\frac {\sqrt{\det\,\biglb(\df,\df\bigrb)_\perp}}
{2\pi\sqrt{\det\,\Delta}}\hbox{\rm Tr}\,\biglb(\df,\df\bigrb)_\perp.
\]
But from (\ref{3.6}) and (\ref{3.7}) and taking into account the 
block-Cholesky decomposition introduced above 
\[\det\,\Delta=\det\,\rmK^2=\det\,\biglb(f,f\bigrb)\ 
\det\,\biglb(\df,\df\bigrb)_\perp.
\]
Thus, we get a simple formula for the average density of 
complex roots extending the Shepp and Vanderbei 
one \cite{SV95} to random Gaussian homogeneous generalized 
polynomials:
\begin{equation}
{\cal D}=\frac{\hbox{\rm Tr}\,\biglb(\df,\df\bigrb)_\perp}
{2\pi\sqrt{\det\,\biglb(f,f\bigrb)}}.
\label{3.12}\end{equation}

Let us now return to the general case of monic polynomials, 
when the deterministic part $\phi$ is not identically zero. 
The joint distribution function of the values of the 
polynomial and its derivative $P$ and $\dot P$ coincides with 
the one for $F$ and $\dot F$ shifted by their deterministic parts 
$\phi$ and, respectively, $\dot\phi$. Thus, for $\Im (z)\ne 0$,
\begin{eqnarray}
{\cal D}=\int\biglb[\tilde{\alpha},\tilde{\alpha}\bigrb]
{\cal P}(-\phi,\tilde{\alpha}-\dot\phi)\rmd^{(2)}\tilde{\alpha},
\label{3.13}\end{eqnarray}
where the function ${\cal P}$ is again given by (\ref{3.5}):
\begin{equation}
{\cal P}(-\phi,{\tilde{\alpha}}-\dot\phi)=
\frac 1{(2\pi)^2\sqrt{\det\,\Delta}}\exp \biggl [
-\frac 12 \big(-\phi^T,{\tilde{\alpha}}^T-\dot\phi^T\big)
\Delta^{-1}\left ({-\phi\atop\tilde{\alpha}-\dot\phi}\right )
\biggr ]. \label{3.14}
\end{equation}
The inhomogeneous Gaussian integral (\ref{3.13}) with $\calP$ 
given by (\ref{3.14}) is calculated in a similar way 
to the previous one by using the block-Cholesky decomposition 
(\ref{3.7}-\ref{3.10}) and introducing a new integration variable
\[\tilde{\beta}=\tilde{\alpha}-\dfi_{\perp},\]
where
\begin{equation}
\dfi_{\perp}=\dfi -\rmT_\perp\phi.
\label{3.15}\end{equation}
Finally, the expected density of roots in the monic case is 
\begin{equation}
{\cal D}=\frac{\exp\left\{-\frac 12\biglb[\phi,\biglb(f,f\bigrb)^{-1}\phi 
\bigrb]\right\}}{2\pi\sqrt{\det\,\biglb(f,f\bigrb)}}\Big\{
\hbox{\rm Tr}\,\biglb(\df,\df\bigrb)_\perp
+\biglb[\dfi _{\perp} ,\dfi _{\perp}\bigrb]\Big\},
\label{3.16}\end{equation}
where $\biglb(\df,\df\bigrb)_\perp$ and $\rmT_\perp$ were defined 
above in (\ref{3.8}) and (\ref{3.9}).

\subsection{Average density of real roots}\label{realr}
On the real axis the situation is rather different. 
There the imaginary parts of all $f_k$ and $\df_k$ 
vanish due to the reality condition (\ref{1.12}). 
The rank of the matrix $\Delta$ is equal to 2 on the 
real axis so that the above calculations are invalid there.

For small values of $y=\Im (z)$ the real and imaginary 
parts of the homogeneous polynomial $F$ are
\[F_1(x,y)=F(x,0)+\calO (y^2),~~F_2(x,y)=yF^\prime (x,0)+\calO (y^3),\]
where $F(x,0)$ is real\footnote{We remind the reader that 
the real and imaginary parts of a complex quantity are denoted 
by the indices $1$, respectively $2$.} and we use the prime to 
denote the derivative with respect to $x$. Introducing this 
into (\ref{1.8}) --- the definition of $\calP$ 
\begin{eqnarray}
\calP (\alpha ,\tilde{\alpha};z)=&\E&
\bigg\{\d\big (\alpha_1-F_1(x,0;\ci )+\calO(y^2)\big )
\d\big (\alpha_2-yF^\prime(x,0;\ci )+\calO (y^3)\big )
\nonumber\\&\times&
\d\big (\tilde{\alpha}_1 -F^\prime(x,0;\ci )+\calO(y^2)\big )
\d\big (\tilde{\alpha}_2-yF^{\prime\prime}(x,0;\ci )+\calO (y^3)\big )\bigg\}, 
\label{3.17}
\end{eqnarray}
and comparing the arguments of the second and third 
deltas we see that the second one may be written as  
$\d\big(\alpha_2-y\tilde{\alpha}_1+\calO (y^3)\big )$. 
Setting the argument $\alpha_2=0$, this becomes
\[|\tilde{\alpha}_1|^{-1}\d(y)+
|y|^{-1}\d\big (\tilde{\alpha}_1-\calO(y^2)\big ). \]
Substituting the second term in the above sum into (\ref{3.17}) 
would yield the small $|y|$ asymptotic behaviour of the density 
of complex roots (\ref{3.12}) which we will obtain a little 
further on in \ref{realasym}. 

The first term, which is zero off the real axis, generates the
singular component of 
$\calP(\alpha_1,0,\tilde{\alpha}_1,\tilde{\alpha}_2;x,y)$ 
\begin{equation}
\calP_{sing}(\alpha_1,0,\tilde{\alpha}_1,\tilde{\alpha}_2;z)=
|\tilde{\alpha}_1|^{-1}\calP_0(\alpha_1,\tilde{\alpha}_1;x)
\d(\tilde{\alpha}_2)\d(y), \label{3.18}
\end{equation}
where $\calP_0$ is the density for the joint distribution 
function of the {\em real} values of $F(x,0)$ and $F^\prime(x,0)$. 

In the Gaussian case $\calP_0$ is also a Gaussian, determined by 
the 2$\times$2 correlation matrix of the (real) values of $F$ 
and $F^\prime$: 
\begin{equation}
\Lambda(x)=\pmatrix{
\biglb(\f(x),\f(x)\bigrb)& \biglb(\f(x),\f^\prime(x)\bigrb)\cr 
\biglb(\f^\prime(x),\f(x)\bigrb)& 
\biglb(\f^\prime(x),\f^\prime(x)\bigrb)
}.\label{3.19}\end{equation}
We use here the notations introduced in the previous section for 
the real n-vector $\f(x)$, which has components $f_{k1}(x,0)$,
and the bold round brackets (\ref{2.6}) for the contraction over 
the Latin indices.

Thus, in the Gaussian case: 
\begin{equation}
{\cal P}_0(\alpha_1,\tilde{\alpha}_1;x)=\frac 1{2\pi\sqrt{\det\,\Lambda(x)}}
\exp \biggl [-\frac 12 (\alpha_1,\tilde{\alpha}_1)
\Lambda^{-1}(x)\left ({\alpha_1 \atop\tilde{\alpha}_1}\right )\biggr ].
\label{3.20}\end{equation}

Substituting the singular component $\calP_{sing}$, with $\calP_0$ 
given by (\ref{3.20}), into (\ref{3.3}) and performing the 
integral leads to Kac's formula \cite{Kac59} for the 
average density of real roots in the homogeneous case
\begin{equation}
\calD_0(x,y)=\d(y)\frac{\sqrt{\biglb(\f(x),\f(x)\bigrb)
\biglb(\f^\prime(x),\f^\prime(x)\bigrb)-
\biglb(\f(x),\f^\prime(x)\bigrb)^2}}{\pi\biglb(\f(x),\f(x)\bigrb)}.
\label{3.21}\end{equation}

Let us now obtain the singular component of the average density 
of roots for monic polynomials. Substituting $\calP_{sing}$, 
(\ref{3.18}-\ref{3.20}), shifted by the deterministic 
part into (\ref{3.3}) and performing the integration yields 
\begin{equation}
\calD_0(x,y)=\d(y)\frac{\sqrt{\det\Lambda(x)}}{\pi\biglb(\f(x),\f(x)\bigrb)}
H\bigl[w(x)\bigr]\exp\left\{{-\frac{\biglb[\phi(x,0),\phi(x,0)\bigrb]}
{2\biglb(\f(x),\f(x)\bigrb)}}\right\},
\label{3.22}\end{equation}
where  
\begin{equation}
H(w)=\exp\left(-\frac{w^2}{2}\right)+\sqrt{\frac{\pi}2}w\ 
\hbox{{\rm erf}}(w), \label{3.23}
\end{equation}
\begin{equation}
w(x)=\sqrt{\frac{\biglb(\f(x),\f(x)\bigrb)}{\det \Lambda(x)}}
\left\vert\phi^\prime(x,0)-\frac{\biglb(\f(x),\f^\prime(x)\bigrb)}
{\biglb(\f(x),\f(x)\bigrb)}\phi(x,0)\right\vert \label{3.24}
\end{equation}
and the error function is
\[\hbox{erf}(s)=\sqrt{\frac 2{\pi}}\int_0^s\rme^{-\frac{t^2}2}\rmd t.\]
The function $H(w)$ is monotonically increasing. $H(0)=1$ and 
$H(w)\approx \sqrt{\pi /2}w$ as $w\to +\infty$. 

For homogeneous polynomials with non-central Gaussian distributions
the density of real roots was obtained by Edelman and Kostlan [6]. 
Taking a singular limit of the covariance matrix in section 5 
of [6] will also yield (\ref{3.22}).

Thus, in the Gaussian case the average density of zeros has 
a regular component given by (\ref{3.16}) in terms of the 
2$\times$2 matrices $\biglb(f,f\bigrb)$, $\biglb(\df ,f\bigrb)$ 
and $\biglb(\df ,\df\bigrb)$. For the real type, (\ref{1.12}), 
polynomials considered in this paper the average density of roots 
has also a singular component, localized on the real line, given 
by the generalized Kac type formula (\ref{3.22}).

\subsection{Asymptotic behavior of the average density of complex 
roots near the real axis}\label{realasym}
Let us now study the asymptotic behavior of the density 
of complex roots near the real axis. For small $y=\Im (z)$, 
the smallest eigenvalue of the matrix 
\[\biglb(f,f\bigrb)\approx\pmatrix{
\biglb(\f ,\f\bigrb)&y\biglb(\f ,\f^\prime\bigrb)\cr
y\biglb(\f ,\f^\prime\bigrb)&y^2\biglb(\f^\prime ,\f^\prime\bigrb)}\]
goes to zero as 
$\left [\biglb(\f^\prime ,\f^\prime\bigrb)-
\biglb(\f ,\f^\prime\bigrb)^2/\biglb(\f ,\f\bigrb)\right ]y^2$. 
The argument of the exponential in the density of complex roots 
(\ref{3.16}) goes to a finite limit; while the other factor 
in the numerator is $\calO (y^2)$. Thus, the average density of 
complex roots goes to zero as $\calO (|y|)$:
\begin{equation}
{\cal D}(x,y)=\frac{|y|\rme^{-{\frac{1}{2}}
\big (\phi(x,0)~ \phi^\prime(x,0)\big )\Lambda(x)
\left ({{\phi(x,0)}\atop{\phi^\prime(x,0)}}\right )}}
{2\pi\sqrt{\det\Lambda(x)}}
\Big\{\biglb[\biglb(\hat{Q}\f,\hat{Q}\f\bigrb)\bigrb]+
\biglb[\hat{Q}\phi ,\hat{Q}\phi \bigrb]\Big\}+\calO(|y^3|), 
\label{3.25}\end{equation}
where the 2$\times$2 matrix $\Lambda (x)$ was defined above in 
(\ref{3.19}) and the linear operator $\hat{Q}$ is defined by
\begin{equation}
\hat{Q}h(x) =h^{\prime\prime}-\frac{
\left[\biglb(\f ,\f\bigrb)\biglb(\f^\prime ,\f^{\prime\prime}\bigrb)-
\biglb(\f,\f^{\prime}\bigrb)\biglb(\f,\f^{\prime\prime}\bigrb)\right]h^\prime
+\left[\biglb(\f^{\prime},\f^{\prime}\bigrb)\biglb(\f,\f^{\prime\prime}\bigrb)
-\biglb(\f ,\f^{\prime}\bigrb)\biglb(\f^\prime ,\f^{\prime\prime}\bigrb)
\right]h}{\det\Lambda}. \label{3.27}
\end{equation}
We omitted all the arguments $(x)$ of the functions 
appearing in the right hand side of (\ref{3.27}).

Thus, the real axis attracts the roots in its vicinity to the 
singular component located on it, depleting the density of 
roots in its neighborhood.

\subsection{High-- and low-disorder limits}\label{hi-lo}
With our definition of the monic random polynomial (\ref{1.11}) 
we may always consider that the random part of the polynomial 
(the coefficients $c_k$ in (\ref{1.11})) has expectation value 
equal to zero. Let us rescale the deterministic part $\phi$
of the polynomial to $\Gamma\phi$, introducing a real 
parameter $\Gamma \ge 0$. The random polynomial is now
\begin{equation}
P(z)=\Gamma\phi (z)+F(z).\label{3.28}
\end{equation}
The form (\ref{3.28}) allows us to interpolate from 
the monic to the homogeneous case. In the Gaussian case 
this is equivalent to considering the coefficients $c_k$ 
of the random polynomial as realizations of a Gaussian process 
with zero average and dispersion 
\begin{equation}
\delta = \Gamma^{-1}, \label{3.29}
\end{equation}
instead of (\ref{3.1}). 

For small values of $\Gamma \to 0$, (the large randomness limit, 
$\delta\to\infty$) the average density of roots $\calD$, (\ref{3.16}), 
approaches the density for homogeneous polynomials, (\ref{3.12}). 
In the large $\Gamma$ limit, equivalent to small randomness, 
$\delta~\to ~0$ in Eq.(\ref{3.29}), the average density of roots 
concentrates near the roots of the deterministic part $\phi(z)$ 
and decays exponentially away from them. Indeed, inspection of 
(\ref{3.16}) shows that at fixed $z=x+\rmi y$, which is not a 
zero of $\phi$, the average density of roots $\calD$, goes to 
zero exponentially when $\Gamma\to\infty$.  

Let now $z_0 =x_0+\rmi y_0$ be a complex zero of $\phi$ having 
multiplicity $k\ge 1$:
\begin{equation} 
\phi(z)=c(z-z_0)^k+\calO\bigl((z-z_0)^{k+1}\bigr), 
\label{3.30}\end{equation}
and assume that the matrix $\Delta(z_0)$ is non-singular. 
Then, to leading order in $\Gamma\to +\infty$ and for small 
values of $\rho =|z-z_0|$, the average density of roots is 
\begin{equation}
\calD(x,y)\approx 
Ak^2\Gamma^2\rho^{2k-2}\rme^{-B(\vp_0)\Gamma^2\rho^{2k}},
\label{3.31}\end{equation}
where 
\begin{equation}
A=\frac{c^2}{2\pi\sqrt{\lambda_1\lambda_2}},
\label{3.32}\end{equation}
\begin{equation}
B(\vp_0)=\frac 12 c^2\bigl[\lambda_1+\lambda_2+
|\lambda_1-\lambda_2|\cos (2k\vp_0 +\chi_0)\bigr],
\label{3.33}\end{equation}
$\vp_0=\arg \bigl(z-z_0\bigr)$,
\begin{equation}
\chi_0=\arctan\frac{2\biglb(f,f\bigrb)_{12}}
{\biglb(f,f\bigrb)_{11}-\biglb(f,f\bigrb)_{22}},
\label{3.34}\end{equation}                     
and $\lambda_1,~\lambda_2$ are the (positive) eigenvalues of 
the 2$\times$2 matrix $\biglb(f,f\bigrb)$. 

Now, it is readily seen that for $\Gamma\to\infty$, 
(\ref{3.31}) goes to $k\delta^{2}(z-z_0)$ in the sense of 
distributions. For large but finite values of $\Gamma$ 
the average density of roots is a Gaussian centered at 
$z_0$ for $k=1$. 

For $k>1$ if the eigenvalues of $\biglb(f,f\bigrb)$ are equal, 
$\lambda_1=\lambda_2$, then the surface $S=\calD(x,y)$ has 
an annular maximum for $\rho^{2k}=(1-k^{-1})/B\Gamma^2$. 
If $\lambda_1\ne\lambda_2$, the annular maximum splits 
into $2k$ individual maxima located at 
$2k\vp_0+\chi_0=(2M+1)\pi,~M=0,\ldots,2k-1$ and 
$\rho^{2k}=(1-k^{-1})/B_{min}\Gamma^2$, where 
$B_{min}=c^2\lambda_{min} = \min_\vp B(\vp)$~ is the 
minimal value of (\ref{3.33}).

\end{section}       

\begin{section}{Asymptotic density of roots for monic 
algebraic polynomials} \label{3}

In this section we investigate the average density of 
roots for algebraic monic polynomials,
\begin{equation}
P_n(z) = z^n + \delta \sn c_kz^{k-1}, \label{4.1}   
\end{equation}
in the large $n$ limit.
Here the parameter $ \delta = \Gamma^{-1} \ge 0$ is the 
dispersion of the original Gaussian distribution which 
was transformed to the normal one,  Eq.(\ref{3.1}), 
as mentioned in the preceding section. 

The roots of the polynomial (\ref{4.1}) coincide with 
the roots of
\begin{equation}
\Gamma P_n(z) = \Gamma z^n + \sn c_kz^{k-1}, \label{4.2}   
\end{equation}
so that we may use the rescaled $\phi$, (\ref{3.28}), as in 
the preceding section. The large $\Gamma$ (low disorder)  
asymptotic obtained there near complex roots of the deterministic 
part $\phi$ cannot be used straightforwardly since in our case 
$\phi $ has a highly degenerate {\em real} zero at the origin.

The calculation of the $2\times 2$ matrices $\biglb(f,f\bigrb)$, 
$\biglb(f,\df \bigrb)$ and $\biglb(\df ,\df \bigrb)$ in this case 
is presented in some detail in Appendix \ref{app-a}. There we show 
that the matrix elements may be expressed in terms of the functions
\begin{eqnarray}
C_n(r^2, 2\vp)& = ~~&\frac{1-r^2\cos 2\vp - r^{2n}\cos 2n\vp + 
r^{2n+2}\cos (2n-2)\vp}{1-2r^2\cos 2\vp + r^4}, \label{4.3} \\
S_n(r^2, 2\vp)& = ~~&\frac{r^2\sin 2\vp - r^{2n}\sin 2n\vp + 
r^{2n+2}\sin (2n-2)\vp}{1-2r^2\cos 2\vp + r^4}, \label{4.4}
\end{eqnarray}
and their derivatives with respect to $r$. Here 
\[z = x + iy = r{\rm e}^{i\vp}. \]

The matrix $\biglb(f,f\bigrb)$ is thus given by
\begin{equation}   
\biglb(f,f\bigrb)=\frac{1}{2}\pmatrix{
\Cnzero + \Cn & \Sn \cr
\Sn & \Cnzero -\Cn \cr}.\label{4.5}
\end{equation}
Its eigenvalues are $\Cnzero \pm \sqrt{C_n^2(r^2,2\vp )+
S_n^2(r^2,2\vp )}$. The smallest is positive for all 
$\vp$ such that $\sin\vp\ne 0$, as shown in \ref{app-a}.

In the Appendix we expressed the matrix elements of 
$\biglb(\df ,f\bigrb)$ and $\biglb(\df ,\df \bigrb)$ in terms 
of derivatives of the matrix $\biglb(f,f\bigrb)$ with respect to $r$. 
We may now rewrite Eqs.(\ref{A10} - \ref{A11}) in matrix form 
using the orthogonal matrix
\begin{equation}   
U(\vp ) = \pmatrix{
\cos\vp & \sin\vp\cr
-\sin\vp & \cos\vp\cr}. \label{4.6}
\end{equation}  
\begin{equation} 
\biglb(\df ,f\bigrb) =\frac{1}{2}U(\vp )
\frac{\partial\biglb(f,f\bigrb)}{\partial r}, 
\label{4.7}
\end{equation}  
\begin{equation} 
\biglb(\df ,\df \bigrb) =\frac{1}{4r}U(\vp )
\frac{\partial }{\partial r}\bigg [
r\frac{\partial\biglb(f,f\bigrb)}{\partial r}\bigg ]U^T(\vp). 
\label{4.8} \end{equation}  
Here $U^T$ is the transposed of $U$.

For examining the behaviour of the density of complex 
roots (\ref{3.16}) in the vicinity of the unit circle, 
let us introduce a new, logarithmically rescaled variable 
\begin{equation} 
\rme^\xi =r^{2n}.\label{4.9}
\end{equation}

Inspection of (\ref{3.16}) and comparison with (\ref{4.7}), 
(\ref{4.8}), (\ref{3.8}), (\ref{3.9}) and (\ref{3.15}) 
shows us that we need to estimate the matrices 
$\biglb(f,f\bigrb)$ and its inverse, the matrix
\[U^T(\vp )\rmT_\perp =\frac{1}{2}
\frac{\partial\biglb(f,f\bigrb)}{\partial r}\biglb(f,f\bigrb)^{-1}, \]
and the second derivative with respect to $r$ of the function $\Cnzero$.
The functions $\Cn$ and $\Sn$, (\ref{4.3}) and (\ref{4.4}), have 
rapid oscillations with $\vp$.

Let us study the quotient 
\[Q=\frac{\Cn}{\Cnzero}=
\frac{\left ( 1-r^{2}\right )
\left (1-r^2\cos 2\vp -r^{2n}\cos 2n\vp +r^{2n+2}\cos (2n-2)\vp\right )}
{\left (1-r^{2n}\right )\left (1-2r^2\cos 2\vp +r^4\right )}, \]
for $|\xi|\ll n$. The absolute value of the sum of terms 
proportional to $r^{2n}$ in the numerator, which are fast 
oscillating with $\vp$, does not exceed 
$r^{2n}\sqrt{1-2r^2\cos 2\vp +r^4}$. Noting that 
$1-2r^2\cos 2\vp +r^4=(1-r^2)^2+(2r\sin\vp )^2$, we obtain 
\[|Q|<\frac{1-r^2}{2r|\sin \vp |}\left (\frac 1{1-r^{2n}}+
\frac 1{r^{-2n}-1}\right )\approx\frac{\xi}{2n|\sin\vp |}
\coth \frac{\xi}{2},\]
where we used the obvious inequalities $2ab<a^2+b^2$ and 
$b<\sqrt{a^2+b^2}$ to estimate the $\vp$ dependent terms. 
A similar estimate can be obtained for the quotient $\Sn /\Cnzero$.

Thus, for 
\begin{equation} 
\tau =\frac{\xi}{2n|\sin \vp|}\coth \frac{\xi}{2}\ll 1,\label{4.10}
\end{equation}
the matrix $\biglb(f,f\bigrb)$ is asymptotically proportional 
to the unit one times $\Cnzero$. Substituting the 
$|\xi |\ll n$ asymptotic behavior of $\Cnzero$, (\ref{A19})
we obtain that $\biglb(f,f\bigrb)$ is equal to $n$ times 
a universal function of $\xi$, which does not depend 
on $\vp$ and $n$:
\begin{equation} 
\biglb(f,f\bigrb)= \frac{n}{2}\ \frac{{\rm e}^\xi - 1}{\xi}
\biggl[1+\calO(\tau )\biggr].  
\label{4.11}\end{equation} 
Noting that (\ref{4.10}) implies $\xi\ll n$, we see that 
we may use (\ref{4.11}) also for estimating the derivatives of 
\biglb(f,f\bigrb). 

Let us first look at the homogeneous case, 
$P_n(z)=\szeron \c_kz^{n-k}$, corresponding to $\Gamma =0$  
and recover the result obtained by Shepp and Vanderbei \cite{SV95}.
Substituting  (\ref{4.11}) into (\ref{4.7}), 
(\ref{4.8}), we obtain after a little algebra on (\ref{3.12})
\begin{equation}
{\cal D}({\rm e}^{\frac{\xi}{2n}},\vp) \approx
-\frac{n^2}{\pi}\frac{\rmd}{\rmd\xi} 
\left (\frac{1}{\xi} - \frac{1}{{\rm e}^\xi - 1}\right ).
\label{4.12}\end{equation} 

Thus, for $\tau\ll 1$, (\ref{4.10}), the expected 
density of complex roots for homogeneous ($\Gamma = 0$) Gaussian 
polynomials is asymptotically equal to $n^2$ times a simple 
universal symmetric function of $\xi$. For large values of $|\xi |\ll n$
the density has inverse power behavior, ${\cal D}\sim\xi^{-2}$. 

In Fig. \ref{fig1} we plot the  renormalized average density 
of complex roots, $n^{-2}D$, for homogeneous real 
Gaussian polynomials of degree $99$ as a function of 
$\xi=2n\ln |z|$ and $\theta=\arg (z)$. We see that at $n=100$ the 
concordance with the asymptotic formula (\ref{4.12}) is good, 
excepting in the vicinity of the real axis  where the condition 
(\ref{4.10}) is invalid.

Performing the same substitutions on (\ref{3.16}) and 
noting that for $\phi=\Gamma z^n$ the $2$-vector $\dfi$ satisfies
\[\dfi = \frac{n}{r}U(\vp )\phi ,\]
we obtain after a little algebra a simple asymptotic formula,  
valid for $\tau\ll 1$, for the average density of complex roots 
${\cal D}$ of monic algebraic polynomials:
\begin{equation}
{\cal D}({\rm e}^{\frac{\xi}{2n}},\vp) = -\frac{n^2}{\pi}\frac{d}{d\xi} 
\biggl\{\left (\frac{1}{\xi} - \frac{1}{{\rm e}^\xi - 1}\right )\exp
\left [{-\frac{\Gamma ^2}{n}\frac{\xi}{1 - {\rm e}^{-\xi}}}\right ]\biggl\}.
\label{4.15}\end{equation} 

In the monic case the expected density of roots is also 
asymptotically equal to $n^2$ times a universal function of
$\xi$ and $\Gamma^2/n$.  For nonzero values of $\Gamma/\sqrt{n}$ 
the large $|\xi|\ll n$ asymptotic behavior of (\ref{4.15}) is 
exponential decay for large positive $\xi$ and remains 
inverse-power for large negative values of $\xi$. 
This asymmetry becomes rather pronounced in the case of 
large values of the parameter $\Gamma^2/n$, which was 
investigated numerically in \cite{FMMB}. 

Let us estimate $n^{-1}\calN_{out}(R)$, the fraction of 
the expected number of roots outside a disk of radius $R$ 
centered at $z = 0$. For large $n$ and $|\ln R|\ll 2n$ 
the total number of real roots is $\calO(\log n)$ and 
since the sectors of angle $\calO(\pi /n)$ near the 
real axis contain a number of roots comparable 
to the error of the asymptotic relation (\ref{4.15}), 
we may also use it there:
\[\calN_{out}(R)=\frac 1n \int_{\xi_R}^\infty\rmd\xi\rme^{\xi/n}
\int_0^\pi\rmd\vp {\cal D}({\rm e}^{\frac{\xi}{2n}},\vp).\]
Here $\xi_R=-\ln R/(2n)$ and we may exploit the exponential decay  
of $\calD$ for large $\xi$ and replace $\rme^{\xi /n}$ by $1$ 
under the integral if $|\xi_R|\ll n$. 

The fraction of roots outside a disk of radius 
$R$ is thus asymptotically equal to
\begin{equation}
\frac 1n\calN_{out}(R)\approx 
\bigg (\frac{1}{2n\ln (R)}-\frac{1}{R^{2n}-1}\bigg )
\exp \bigg (-\frac{2\Gamma ^2\ln (R)}{1 - R^{-2n}}\bigg ).
\label{4.16}\end{equation} 
For $R\to 0$ this goes to $n$.

In Fig. \ref{fig2} we plot the (re)normalized density of roots 
for $n=10$ and $N=\Gamma^2=100$.  It has nine sharp peaks in
each of the half-planes $\Im (z){{>}\atop{<}}0$. Our analysis 
n \ref{hi-lo} predicts the splitting of a {\em complex} 
$n$ times degenerate zero of the deterministic part into 
$2n$ maxima. In the present case we have only $2n-2$ peaks 
because the zero of the deterministic part is {\em real} 
and the other two peaks of the distribution are on the singular 
component. An interesting feature of this splitting is the 
fact that there are twice as many peaks as there are roots. 
Thus, on a typical realization of the random process we expect 
to find the $n$ roots of the polynomial located near 
{\em half} of the positions of the maxima. 

At $n=30$ and $N=1024$ there is still some oscillation
in the expected number of roots in an angular sector 
as can be seen in Fig. \ref{fig3}. 
For larger values of $n$, the density of complex roots 
approaches the $\vp$ independent asymptotic form (\ref{4.15}).
In Fig. \ref{fig4} this convergence is shown for
$\frac{2\pi}{5}<\vp <\frac{\pi}{2}$ 
at $N = \Gamma^2=10n$. The number of extrema increases 
with $n$ while their values become closer. The first 
several maxima near the real axis have a larger amplitude
than those in the domain of universality. Since 
these are at a distance $\sim n^{-1}$ from the real axis, 
the linear with $|\Im (z)|$ fall to zero of the 
density of roots may become rather steep as can be seen 
in Fig. \ref{fig5}.
\end{section}


\appendix
\begin{section}{Matrix elements -- algebraic case}
\label{app-a}

The density of complex roots is given by Eq.(\ref{3.16}) in terms 
of the 2$\times$2 matrices $\biglb(f,f\bigrb)$, $(f,\df )$ and 
$\biglb(\df ,\df\bigrb)$ defined in Section \ref{not}.  Let us 
calculate them explicitly in the case when the polynomials are 
algebraic, {\it i.e.} when the basis functions are given by:
\begin{equation}
f_k(z)= z^{k-1},~~~k=1,\dots ,n. \label{A1}
\end{equation}   
Taking $z = r{\rm e}^{i\vp}$, the components of the 
$2n$-vectors $f$ and $\df$ are 
\begin{eqnarray}  
f_{k1}&=r&^{k-1}\cos (k-1)\vp, \nonumber \\ 
f_{k2}&=r&^{k-1}\sin (k-1)\vp , \label{A2}\\
\df_{k1}&=(&k-1)r^{k-2}\cos (k-2)\vp, \nonumber \\ 
\df_{k2}&=(&k-1)r^{k-2}\sin (k-2)\vp  \label{A3}
\end{eqnarray}

Let us start with the 2$\times$2 real symmetric matrix $\biglb(f,f\bigrb)$.
\begin{eqnarray}  
\biglb(f,f\bigrb)_{11}& =&{\szeron} r^{2k} \cos^2 k\vp = 
\frac{1}{2} \szeron r^{2k}\big (1 + \cos 2k\vp \big)  \nonumber \\
~~& =&\frac{1}{2}\bigl [ \Cnzero + \Cn \bigr ]. \label{A4}
\end{eqnarray}
Here, we defined the function $C_n(x,\chi )$ as the real 
part of the sum
\begin{eqnarray}
\szeron x^k {\rm e}^{ik\chi }&=& C_n(x,\chi) + iS_n(x,\chi), \label{A5} \\
C_n(x,\chi)&=&\frac{1-x\cos\chi - x^n\cos n\chi + x^{n+1}\cos (n-1)\chi}
{1-2x\cos\chi + x^2}, \label{A6} \\
S_n(x,\chi)&=&\frac{x\sin\chi - x^n\sin n\chi + x^{n+1}\sin (n-1)\chi}
{1-2x\cos\chi + x^2}. \label{A7}
\end{eqnarray}

In a similar way, the other elements of the symmetric matrix 
$\biglb(f,f\bigrb)$ are given by
\begin{eqnarray} 
\biglb(f,f\bigrb)_{12}&= \frac{1}{2}&\Sn, \nonumber \\
\biglb(f,f\bigrb)_{22}&= \frac{1}{2}&\bigl [ \Cnzero - \Cn \bigr ], 
\label{A8}\end{eqnarray} 
where the function $S_n(x,\chi)$ is defined in (\ref{A7}). 
The determinant of the matrix $\biglb(f,f\bigrb)$ is 
\begin{eqnarray}
\det\,\biglb(f,f\bigrb)&= \frac{1}{4}&\bigl [ C_n^2(r^2,0)-
C_n^2(r^2,2\vp )-S_n^2(r^2,2\vp ) \bigr ] \nonumber \\
&= \frac{1}{4}&\Biggl [ \biggl(\frac{1-r^{2n}}{1-r^2}\biggr)^2 - 
\frac{1-2r^{2n}\cos 2n\vp +r^{4n}}{1-2r^2\cos 2\vp +r^4} \biggl ].
\label{A9}
\end{eqnarray}            
It is readily seen that the determinant is greater than zero if 
$\sin \vp\ne 0$ {\it i.e. off the real axis.}

The sums appearing in the definitions of $\biglb(f,\df\bigrb)$ 
and $\biglb(\df ,\df\bigrb)$ are similar. The factors $k$ and 
$k^2$ may be dealt with using the relation 
$\frac{dx^k}{dx} = k x^{k-1}$. Thus, the elements of the matrices 
$\biglb(f,\df\bigrb)$ and $\biglb(\df ,\df\bigrb)$ can be expressed 
in terms of derivatives with respect to the first argument of the 
functions $C_n$ and $S_n$ defined by (\ref{A6}) and (\ref{A7}) above.
\begin{eqnarray} 
\biglb(\df,f\bigrb)_{11}&=~&{\szeron}kr^{2k-1} \cos k\vp \cos (k-1)\vp = 
{\szeron}kr^{2k-1} \big [\cos\vp \cos^2k\vp +\sin\vp \sin k\vp \cos k\vp
\big ]  \nonumber\\
&=~&\frac{1}{2} \frac{\partial }{\partial r} \Big[\cos \vp 
\biglb(f,f\bigrb)_{11}+\sin\vp \biglb(f,f\bigrb)_{21}\Big], \nonumber\\
\biglb(\df,f\bigrb)_{12}&=~&\frac{1}{2} \frac{\partial }{\partial r} 
\Big[\cos\vp\biglb(f,f\bigrb)_{12} +\sin\vp\biglb(f,f\bigrb)_{22}\Big], 
\label{A10}\\
\biglb(\df,f\bigrb)_{21}&=~&\frac{1}{2} \frac{\partial }{\partial r} 
\Big[\cos\vp\biglb(f,f\bigrb)_{21} -\sin\vp\biglb(f,f\bigrb)_{11}\Big], 
\nonumber\\
\biglb(\df,f\bigrb)_{22}&=~&\frac{1}{2} \frac{\partial }{\partial r} 
\Big[\cos \vp \biglb(f,f\bigrb)_{22} - \sin\vp\biglb(f,f\bigrb)_{12}\Big]. 
\nonumber
\end{eqnarray} 
\begin{eqnarray}
\biglb(\df ,\df\bigrb)_{11}&=~& {\szeron} k^2r^{2k-2} \cos^2(k-1)\vp
\nonumber\\
&=~&{\szeron} k^2r^{2k-2}\Big [\cos^2\vp\cos^2k\vp +2\sin\vp\cos\vp
\cos k\vp\sin k\vp +\sin^2\vp\sin^2k\vp\Big] \nonumber\\
&=~&\frac{1}{4r} \frac{\partial }{\partial r}r \frac{\partial }{\partial r} 
\Big [\cos^2\vp\biglb(f,f\bigrb)_{11}+ 2\sin\vp\cos\vp \biglb(f,f\bigrb)_{12} 
+ \sin^2\vp \biglb(f,f\bigrb)_{22}\Big ], \nonumber\\
\biglb(\df ,\df\bigrb)_{12}&=~&\frac{1}{4r} \frac{\partial }{\partial r}r 
\frac{\partial}{\partial r} 
\Big [(\cos^2\vp-\sin^2\vp)\biglb(f,f\bigrb)_{12} +
\sin\vp\cos\vp\Big(\biglb(f,f\bigrb)_{22}-\biglb(f,f\bigrb)_{11} 
\Big)\Big],\label{A11}\\
\biglb(\df ,\df\bigrb)_{22}&=~&\frac{1}{4r} \frac{\partial }{\partial r}r 
\frac{\partial }{\partial r} 
\Big [\cos^2\vp\biglb(f,f\bigrb)_{22}- 2\sin\vp\cos\vp \biglb(f,f\bigrb)_{12} 
+ \sin^2\vp \biglb(f,f\bigrb)_{11}\Big ]. \nonumber\\
\end{eqnarray} 

After expressing all the relevant matrix elements in terms of the 
functions $C_n$ and $S_n$, let us now consider their asymptotic 
behavior when $n\gg 1 $. The sum in Eq.(\ref{A5}) and its 
derivatives with respect to $x$ are truncated power series in $x$ 
with power bounded coefficients. Their radius of convergence is $1$. 
For $x<1$ the leading term of its large $n$ asymptotic expansion is 
given by the $n\to\infty$ limit of Eq.(\ref{A5}) yielding a result 
$n$-independent result with exponentially small corrections:
\begin{eqnarray}
C_n(x,\chi)& \approx &\frac{1-x\cos\chi }{1-2x\cos\chi + x^2}
+\calO (x^{n}), \label{A12} \\
S_n(x,\chi)& \approx &\frac{x\sin\chi }{1-2x\cos\chi + x^2}
+\calO (x^{n}). \label{A13}
\end{eqnarray} 

If $x>1$ the series the sum Eq.(\ref{A5}) diverges as $n\to\infty$. 
The leading terms for $C_n$ and $S_n$ are now exponentially large 
in absolute value and rapidly oscillating functions of $\chi$. 
The corrections are again rather small:
\begin{eqnarray}  
C_n(x,\chi)& \approx &x^n\Bigg [ \frac{ x\cos (n-1)\chi - \cos n\chi}
{1-2x\cos\chi + x^2}+\calO (x^{-n})\Bigg ], \label{A14} \\
S_n(x,\chi)& \approx &x^n\Bigg [ \frac{x\sin (n-1)\chi - \sin n\chi}
{1-2x\cos\chi + x^2}+\calO (x^{-n})\Bigg ]. \label{A15}
\end{eqnarray}

Thus, in the large $n$ limit the asymptotic behavior of the 
functions $C_n$ and $S_n$ changes dramatically in a narrow 
annulus of width $\calO (n^{-1})$ near the unit circle, crossing 
over from the behavior Eq.(\ref{A12}) to Eq.(\ref{A14}) 
( respectively from  Eq.(\ref{A13}) to Eq.(\ref{A15}) ).   

For investigating the behavior in the crossover region, 
let us go to a logarithmic scale for the variable $x$:
\begin{equation}
x=\exp \Big (\frac{\xi}{n} \Big ). \label{A16}
\end{equation}
For $\vert\xi\vert \ll n\chi$, the leading asymptotic terms are  
\begin{eqnarray}  
C_n\big({\rm e}^{\frac{\xi}{n}},\chi\big)& \approx &\frac{1}{2}
\Big [1+\frac{\sin (n-\frac{1}{2})\chi}{\sin\frac{\chi}{2}}
{\rm e}^{\xi}\Big ]+\calO\Big (\frac{\xi}{n\chi}\Big ),\label{A17}\\
S_n\big({\rm e}^{\frac{\xi}{n}},\chi\big)& \approx &\frac{1}{2}
\Big [\cot \frac{\chi}{2}-\frac{\cos (n-\frac{1}{2})\chi}{\sin\frac{\chi}{2}}
{\rm e}^{\xi}\Big ]+\calO\Big (\frac{\xi}{n\chi}\Big ).\label{A18}
\end{eqnarray}

Inspection of Eqs.(\ref{A17}-\ref{A18}) shows that the corrections 
are small outside a neighborhood of the point $x=1,~\chi = 0$. 
For $\chi = 0$ and $\vert\xi\vert \ll n$, the leading term 
is\footnote{We will not write here the bulkier expressions which 
are valid for $\vert\xi\vert \ll n$ without\hfill\break 
restrictions on $\chi$.} 
\begin{equation}
C_n\big({\rm e}^{\frac{\xi}{n}},0\big) \approx 
n\frac{{\rm e}^{\xi}-1}{\xi}+\calO (1).\label{A19}
\end{equation}

The asymptotic expansions given by Eqs.(\ref{A12} - \ref{A15}), 
(\ref{A17} - \ref{A19}) may be now used to obtain the large $n$ 
asymptotic behavior of the matrices $\biglb(f,f\bigrb)$, 
$\biglb(f,\df\bigrb)$ and $\biglb(\df ,\df\bigrb)$ in the domains 
$\vert z^n\vert \ll 1$, ~$\vert z^n\vert \gg 1$ and the thin annulus 
which separates them.

\end{section}

\newpage

\begin{figure}
\caption{The renormalized density of complex roots, 
$n^{-2}D(z)$,  for the homogeneous Gaussian polynomial 
$c_1z^{99}+ c_2z^{98}+ \ldots +c_{100}$, $n=100$,  
as a function of $\xi = 2n\ln |z|$ and $\vp = \arg (z)$. 
(a) exact, Eq. (\protect{\ref{3.12}}); (b) asymptotic formula, 
Eq. (\protect{\ref{4.12}}); (c)  contour plots: full line -- exact, 
dotted line -- asymptotic.
\label{fig1} }
\end{figure}

\begin{figure}
\caption{Renormalized density of complex roots for the 
polynomial $ N^{\frac 12}z^{10}+ c_1z^9 + \ldots +c_{10}$, 
$n=10$,  and $N=\Gamma^2=100$, as function of $\xi$ and $\vp$.
\label{fig2} }
\end{figure}

\begin{figure}
\caption{Average number of roots in angular
sectors  for $n=30$ and $N=1024$. Full line -- 
histogram of the averages for 1000 polynomials; 
dashed line -- angular density of roots, 
$\int_0^\infty \rmd r r D(r,\vp)$, obtained by 
numerical integration of Eq. (\protect{\ref{3.16}}).\label{fig3} }
\end{figure}

\begin{figure}
\caption{Renormalized density of roots for 
$N=\Gamma^2=10n$: (a) $n=100$, exact; (b) $n=200$, exact; 
(c) asymptotic formula (\protect{\ref{4.15}}) for $N/n=10$. 
\label{fig4} }
\end{figure}

\begin{figure}
\caption{Detail of the asymptotic behavior of the 
expected density of roots near the real axis for 
$n=30$ and $N=1000$. \label{fig5}}
\end{figure}

\end{document}